\documentclass[11pt]{article} 
\newcommand{\be}{\begin{equation}}
\newcommand{\ee}{\end{equation}}
\newcommand{\bea}{\begin{eqnarray}}
\newcommand{\eea}{\end{eqnarray}}
\newcommand{\sn}{{\rm sn}}
\newcommand{\ds}{{\rm ds}}
\newcommand{\cs}{{\rm cs}}
\newcommand{\ns}{{\rm ns}}
\newcommand{\dn}{{\rm dn}}
\newcommand{\cn}{{\rm cn}}
\newcommand{\sech}{{\rm sech}}

\begin{document}
\vspace{.5in} 
\begin{center} 
{\LARGE{\bf Solutions of Several Coupled Discrete Models 
in terms of Lam\'e Polynomials of Order One and Two}}
\end{center} 

\vspace{.3in}
\begin{center} 
{\LARGE{\bf Avinash Khare*}} \\ 
{Institute of Physics, Bhubaneswar, Orissa 751005, India}
\end{center} 

\begin{center} 
{\LARGE{\bf Avadh Saxena}} \\ 
{Theoretical Division and Center for Nonlinear Studies, Los
Alamos National Laboratory, Los Alamos, NM 87545, USA}
\end{center} 

\vspace{.9in}
{\bf {Abstract:}}  

Coupled discrete models abound in several areas of physics. Here we provide 
an extensive set of exact quasiperiodic solutions of a 
number of coupled discrete 
models in terms of Lam\'e polynomials of order one and two. Some of the models
discussed are (i) coupled Salerno model, (ii) coupled Ablowitz-Ladik model,
(iii) coupled saturated discrete nonlinear Schr\'odinger equation, (iv) coupled 
$\phi^4$  model, and (v) coupled $\phi^6$ model. Furthermore, we show that
most of these coupled models in fact also possess an even broader class of 
exact solutions.\\

\vspace{1.5in}
*Address From January 31, 2011: IISER Pune, India 411021.
\newpage 
  
\section{Introduction} 
In a recent paper \cite{ak} we have obtained solutions of a coupled
$\phi^4$ and and a coupled $\phi^6$ model in terms of Lam\'e polynomials
of order one and two even though most of these solutions are not the solutions 
of the corresponding uncoupled problem. The purpose of the present paper is 
to carry out a similar study for a number of {\it coupled discrete} field theory 
models. In particular, we obtain exact solutions of a (i) coupled Ablowitz-Ladik 
(AL) model, (ii) coupled Salerno model, (iii) coupled saturated discrete nonlinear 
Schr\'odinger equation (DNLSE), (iv) coupled $\phi^6$ model, and (v) coupled
$\phi^4$ model. We also show that unlike the continuum field theory models, 
many of the discrete coupled field theory models possess an even broader class 
of exact solutions. Moreover, we show that as in the uncoupled 
case \cite{krsss}, 
even the coupled AL, coupled Salerno and coupled DNLSE models follow from
the {\it same} Hamiltonian but with a different Poisson bracket (PB) structure.

The motivation for this work comes from the fact that there are many physical 
situations where a discrete field theory is appropriate to model the phenomena 
of interest with a specific coupling between the two fields.  One such 
phenomenon of current intense interest is the coexistence of magnetism and 
ferroelectricity (i.e. magnetoelectricity) in a given material.  This 
is a highly desired functionality in technological applications involving 
cross-field response, switching and actuation.  In general, this 
phenomenon is referred to as multiferroic behavior \cite{multif}.  
Recently, two different classes of (single phase) multiferroics, namely 
the orthorhombically distorted perovskites \cite{kimura} and rare earth 
hexagonal structures \cite{fiebig}, have emerged.  The latter show 
magnetic domain walls in the basal planes which can be modeled by a 
coupled $\phi^4$ model \cite{curnoe} in the presence of a magnetic 
field.  Coupled $\phi^4$ models \cite{aubry,abel,jcp} also arise in the 
context of many ferroelectric and other second order phase transitions.  
The coupled $\phi^4$ model for multiferroics \cite{curnoe} has a 
biquadratic coupling whereas the coupled $\phi^4$ model for a surface 
phase transition with hydration forces \cite{jcp}, relevant in biophysics 
context, has a bilinear coupling.  Other types of couplings are also 
known for structural phase transitions with  strain \cite{das}.  

Examples of coupled discrete AL, coupled discrete Salerno and coupled 
saturated DNLS models are also known \cite{cuevas,rothos,panosbook}.  
Similarly, there are analogous coupled models in field theory \cite{raja,pla}.  
Several related models have been discussed in the literature and their soliton 
solutions have been found \cite{lai,rao,wang,huang,bazeia,zhu,lou,cao} 
including periodic ones \cite{li,liu,llw}.  

The paper is organized as follows. In Sec. II we first show that the coupled
AL, coupled Salerno and coupled DNLSE models can all be obtained from the
same Hamiltonian but with different PB structure. We also obtain additional 
conserved quantities in these models. In Sec. III we provide the solutions 
for the coupled Salerno model in terms of Lam\'e polynomials of order
one and two as well as a broader class of solutions. In Sec. IV we provide 
similar solutions for the coupled AL model. In Sec. V we show that unlike the 
coupled Salerno and coupled AL cases, the coupled saturated DNLSE 
model only admits Lam\'e polynomial \cite{lame} solutions of order one but 
not of order two. Besides, we have not been able to obtain a broader class 
of solutions in this case. Section VI is devoted to solutions of a coupled 
discrete 
$\phi^6$ model \cite{ak,ak2} in terms of Lam\'e polynomials of order one and 
two and also a broader class of solutions. In Sec. VII we discuss solutions 
of a 
coupled $\phi^4$ model introduced by us recently \cite{ak1} in terms of Lam\'e 
polynomials of order two as well as a broader class of solutions.  Note that the
solutions in terms of Lam\'e polynomials of order one have already been 
obtained by us in \cite{ak1}.   Section VIII contains the summary of main 
results and possible future directions.

\section{The Model for Coupled saturated DNLSE, coupled AL and coupled 
Salerno Equations}

We have previously shown \cite{krsss} that the uncoupled Salerno model
\cite{sal}, the uncoupled AL model \cite{al1} and the uncoupled saturated 
DNLSE model can all be deduced from the same Hamiltonian
\be\label{m1}
H=\sum_{n=1}^{N} \bigg [|u_n-u_{n+1}|^2-\frac{\nu}{\mu}|u_n|^2
+\frac{\nu}{\mu^2} \ln(1+\mu|u_n|^2) \bigg ]\,,
\ee
but with different PB structure. We now show that even the {\it coupled} Salerno, 
{\it coupled} AL and the {\it coupled saturated} DNLSE models can all be derived 
from the same Hamiltonian given by 
\bea\label{m2}
&&H=\sum_{n=1}^{N} \bigg [|u_n-u_{n+1}|^2+|v_n-v_{n+1}|^2
-\frac{\nu_1}{\mu_1}|u_n|^2 \nonumber \\
&&-\frac{\nu_2}{\mu_2} |v_n|^2
+\frac{\nu_1}{\mu_1^2} \ln(1+\mu_1|u_n|^2+\mu_2|v_n|^2) \bigg ]\,,
\eea
with the equations of motion in the two field variables $u_n$ and $v_n$ in all 
three cases being
\be\label{m3}
i\dot{u}_n=[u_n,H]\,,~~i\dot{v}_n=[v_n,H]\,.
\ee
The difference in the equations of motion comes from a different definition 
of the PB and consequently a different definition of the time derivative.
The PB structure in all three cases can be compactly written as 
\bea\label{m4}
[U,V]=\sum_{n=1}^{N} \bigg [\frac{\partial U}{\partial u_n}\frac{\partial V}
{\partial u_n^{*}}- \frac{\partial U}{\partial u_n^{*}}\frac{\partial V}
{\partial u_n} 
+ \frac{\partial U}{\partial v_n}\frac{\partial V}
{\partial v_n^{*}}- \frac{\partial U}{\partial v_n^{*}}\frac{\partial V}
{\partial v_n} \bigg ][1+\lambda_1 |u_n|^2+\lambda_2 |v_n|^2]\,.
\eea

\noindent{\bf Coupled saturated DNLSE}

On using Eqs. (\ref{m2}) to (\ref{m4}) with $\lambda_1=\lambda_2
=0$ yields the coupled saturated DNLS equations
\be\label{m5}
idu_n/dt +[u_{n+1}+u_{n-1}-2u_n]+ \frac{\nu_1(\mu_1 \mid u_n \mid^2 
+\mu_2 \mid v_n \mid^2) u_n}
{\mu_1(1+\mu_1 |u_n|^2+\mu_2|v_n|^2)}=0\,, 
\ee
\be\label{m6}
idv_n/dt +[v_{n+1}+v_{n-1}-2v_n] 
+ \frac{\left(\nu_2-\frac{\nu_1 \mu_2^2}{\mu_1^2}\right)v_n +
\nu_2(\mu_1 \mid u_n \mid^2 
+\mu_2 \mid v_n \mid^2) v_n}
{\mu_2(1+\mu_1 |u_n|^2+\mu_2|v_n|^2)}=0\,. 
\ee

\noindent It is easily checked that in this case, apart from the Hamiltonian
(\ref{m1}), two other conserved quantities are power $P_u$ and $P_v$ 
defined by
\be\label{m7}
P_{u} =\sum_{n=1}^{N} |u_n|^2\,,~~P_{v} =\sum_{n=1}^{N} |v_n|^2\,.
\ee

\noindent{\bf Coupled Salerno Model}
 
If instead, we use Eqs. (\ref{m2}) to (\ref{m4}) with $\lambda_1=\mu_1$
and $\lambda_2=\mu_2$ then we obtain the coupled Salerno model with 
field equations
\be\label{m8}
idu_n/dt +[u_{n+1}+u_{n-1}-2u_n]+(\mu_1 \mid u_n \mid^2+\mu_2\mid
v_n \mid^2)\left[u_{n+1}+u_{n-1}+\frac{\nu_1-2\mu_1}{\mu_1}u_n\right] =0\,,
\ee
\be\label{m9}
idv_n/dt +\left[v_{n+1}+v_{n-1}-\left(2+\frac{\nu_1 \mu_2}{\mu_1^2}
-\frac{\nu_2}{\mu_2}\right)v_n\right]+(\mu_1 \mid u_n \mid^2+\mu_2\mid
v_n \mid^2)\left[v_{n+1}+v_{n-1}+\frac{\nu_2-2\mu_2}{\mu_2}v_n\right] =0\,.
\ee

\noindent It is easily checked that in this case, apart from the Hamiltonian
(\ref{m1}), the other conserved quantity is power $P$ given by
\be\label{m10}
P =\sum_{n=1}^{N} \ln[1+\mu_1|u_n|^2+\mu_2 |v_n|^2]\,.
\ee

\noindent{\bf Coupled AL Model}

In the special case when $\nu_1=2\mu_1$ and $\nu_2=2\mu_2$, then the coupled
Salerno model reduces to the coupled AL model with the field equations
\be\label{m11}
idu_n/dt +[u_{n+1}+u_{n-1}-2u_n]+(\mu_1 \mid u_n \mid^2+\mu_2\mid
v_n \mid^2)[u_{n+1}+u_{n-1}] =0\,,
\ee
\be\label{m12}
idv_n/dt +\left[v_{n+1}+v_{n-1}-\frac{2\mu_2}{\mu_1}v_n\right]
+(\mu_1 \mid u_n \mid^2+\mu_2\mid
v_n \mid^2)[v_{n+1}+v_{n-1}] =0\,.
\ee

\noindent It is interesting to note that in this case, apart from the 
Hamiltonian 
(\ref{m1}) and power $P$ as given by Eq. (\ref{m10}), generalized momentum
$P_m$ given by
\be\label{m13}
P_m=\sum_{n=1}^{N} i[\mu_1(u_n u_{n+1}^{*}-u_n^{*} u_{n+1})+
\mu_2(v_n v_{n+1}^{*}-v_n^{*} v_{n+1})]\,,
\ee
is also conserved. 

One remark is in order here. Just as the uncoupled Salerno model interpolates
between AL and DNLSE, it is easy to see that the coupled Salerno model as 
given by Eqs. (\ref{m8}) and (\ref{m9}) also interpolates between coupled AL 
model (as given by Eqs. (\ref{m11}) and (\ref{m12})) and coupled DNLSE.  In 
particular, in the limit $\nu_1=2\mu_1$ and $\nu_2=2\mu_2$, the coupled 
Salerno model Eqs. (\ref{m8}) and (\ref{m9}) go over to the coupled AL model 
Eqs. (\ref{m11}) and (\ref{m12}).  On the other hand, in the limit $\mu_1=\mu_2=0$ 
but with $\mu_2/\mu_1=c$
and $\nu_2 \mu_1^2 = \nu_1 \mu_2^2$ the coupled Salerno model Eqs. (\ref{m8})
and (\ref{m9}) go over to the coupled DNLS equations  
\be\label{m8a}
idu_n/dt +[u_{n+1}+u_{n-1}-2u_n]+\nu_1 (\mid u_n \mid^2+c \mid
v_n \mid^2)u_n =0\,,
\ee
\be\label{m9a}
idv_n/dt +[v_{n+1}+v_{n-1}-2v_n]
+ \nu_1 c(\mid u_n \mid^2+c\mid v_n \mid^2)v_n =0\,.
\ee

\section{Solutions of the Coupled Salerno Model}
  
We now show that the coupled Salerno model as given by Eqs. (\ref{m8}) and
(\ref{m9}) has Lam\'e polynomial solutions 
of order one as well as of order two. In fact, it turns out that the coupled 
model has an even broader class of exact solutions of which Lam\'e polynomial
solutions of order one and two are the special cases. We remind 
that so far as we are aware of, the uncoupled Salerno model has {\it no} 
known exact solutions. 

We start with the ansatz
\be\label{m14}
u_n=f_n \exp{[-i(\omega_1 t+\delta_1)]}\,,
~~v_n=g_n \exp{[-i(\omega_2 t+\delta_2)]}\,,
\ee
with $f_n$ and $g_n$ satisfying
\be\label{m15}
f_n^2+ag_n^2=b\,,~~a,b>0\,.
\ee
Here $a,b>0$ are two positive numbers while $\delta_1$, $\delta_2$ are two 
arbitrary parameters. 
On substituting this ansatz in Eqs. (\ref{m8}) and (\ref{m9}) we find that
this is a consistent ansatz provided
\be\label{m16}
a=\frac{\mu_2}{\mu_1}\,,~~ b =-\frac{1}{\mu_1}\,,~~\omega_1 = \frac{\nu_1}{\mu_1}\,,
~~\omega_2= \frac{\nu_1 \mu_2}{\mu_1^2}\,.
\ee
This implies that such solutions are possible only if $\mu_1,\mu_2<0$. 
Further, since 
\be
\frac{\omega_1}{\omega_2} = \frac{\mu_1}{\mu_2}\,,
\ee
which is a real number, hence the solutions obtained from here are in general 
only quasiperiodic. Only if $\frac{\mu_1}{\mu_2}$ is a rational number, will
the solutions be periodic. 

Clearly this is a very general ansatz which admits a broad class of solutions
including Lam\'e polynomials of order one and two. 

\noindent{\bf Lam\'e polynomial solutions of order one}

(i) One solution is

\be\label{m17}
f_n=A \dn[\beta(n+c_2),m]\,,~~ g_n=B \sqrt{m} \sn[\beta(n+c_2),m]\,,
\ee
provided Eq. (\ref{m16}) is satisfied and further
\be\label{m18}
b=A^2\,,~~\mu_1 A^2 = \mu_2 B^2\,.
\ee
Note that $\beta$ is completely arbitrary. Using the fact that $\dn(x,m)$
has period $2K(m)$ while $\cn(x,m)$ and $\sn(x,m)$ are periodic functions
with period $4K(m)$, it follows that for the solution (\ref{m17}), $u_n,v_n$
satisfy the boundary condition
\be\label{m17a}
u_{n+\frac{2K(m)}{\beta}}= u_n\,,~~v_{n+\frac{4K(m)}{\beta}}=v_n\,.
\ee
Here $K(m)$ is the complete integral of the first kind.

(ii) Another solution is

\be\label{m19}
f_n=A \sqrt{m} \cn[\beta(n+c_2),m]\,,~~ g_n=B \sqrt{m} \sn[\beta(n+c_2),m]\,,
\ee
provided Eq. (\ref{m16}) is satisfied and further
\be\label{m20}
b=m A^2\,,~~\mu_1 A^2 = \mu_2 B^2\,.
\ee
For the solution (\ref{m19}), $u_n,v_n$
satisfy the boundary condition
\be\label{m19a}
u_{n+\frac{4K(m)}{\beta}}= u_n\,,~~v_{n+\frac{4K(m)}{\beta}}=v_n\,.
\ee

In the limit $m=1$, both these solutions go over to the hyperbolic solution
\be\label{m21}
f_n=A \sech[\beta(n+c_2)]\,,~~ g_n=B \tanh[\beta(n+c_2)]\,,
\ee

\noindent{\bf Lam\'e polynomial solutions of order two}

(iii) One solution is
\be\label{m22}
f_n=A \dn^2[\beta(n+c_2),m]+B\,,~~ 
g_n=F \sqrt{m} \sn[\beta(n+c_2),m] \dn[\beta(n+c_2),m]\,,
\ee
provided Eq. (\ref{m16}) is satisfied and further
\be\label{m23}
b=\frac{A^2}{4}\,, ~~\mu_1 A^2 = \mu_2 F^2\,,~~A=-2B\,.
\ee
For the solution (\ref{m22}), $u_n,v_n$ satisfy the boundary condition
(\ref{m17a}).

(iv) Another solution is

\be\label{m24}
f_n=A \dn^2[\beta(n+c_2),m]+B\,,~~ 
g_n=F m \sn[\beta(n+c_2),m] \cn[\beta(n+c_2),m]\,,
\ee
provided Eq. (\ref{m16}) is satisfied and further
\be\label{m25}
b=\frac{m^2 A^2}{4}\,, ~~\mu_1 A^2 = \mu_2 F^2\,,~~(2-m)A=-2B\,.
\ee
For the solution (\ref{m24}), $u_n,v_n$ satisfy the boundary condition
\be\label{m25z}
u_{n+\frac{2K(m)}{\beta}}= u_n\,,~~v_{n+\frac{2K(m)}{\beta}}=v_n\,.
\ee

In the limit $m=1$, both solutions (\ref{m22}) and (\ref{m24}) go over to the
hyperbolic solution
\be\label{m26c}
f_n=A \sech^2[\beta(n+c_2)]+B\,,~~ 
g_n=F \tanh[\beta(n+c_2)] \sech[\beta(n+c_2)]\,. 
\ee

(v) Apart from these, several other solutions are possible. For example
one can have nonperiodic solutions like
\be\label{m26a}
f_n = \frac{A}{\sqrt{1+n^2}}\,,~~g_n=\frac{Bn}{\sqrt{1+n^2}}\,,
\ee
provided Eq. (\ref{m16}) is satisfied and further
\be\label{m25a}
b=A^2\,, ~~\mu_1 A^2 = \mu_2 B^2\,.
\ee
One can obviously write down a wider class of such solutions. For example 
\be\label{m26aa}
f_n = \frac{A\sqrt{1+n^2}}{\sqrt{1+n^2+n^4}}\,,
~~g_n=\frac{Bn^2}{\sqrt{1+n^2+n^4}}\,,
\ee
provided Eqs. (\ref{m16}) and (\ref{m25a}) are satisfied.

(vi) Yet another possible periodic solution is
\be\label{m25b}
f_n=A \cos[\beta(n+c_2)]\,,~~ g_n=B \sin[\beta(n+c_2)]\,,
\ee
provided Eq. (\ref{m16}) is satisfied and further 
\be\label{m25c}
b=A^2\,, ~~\mu_1 A^2 = \mu_2 B^2\,.
\ee
In this case both $u_n,v_n$ satisfy the periodicity condition 
\be\label{m25cz}
u_{n+\frac{2\pi}{\beta}}=u_n\,,~~v_{n+\frac{2\pi}{\beta}}=v_n\,.
\ee

It turns out that apart from the general solution as given by 
Eqs. (\ref{m14}) and (\ref{m15}), there is another possible general
solution given by
\be\label{m26}
u_n=f_n \exp{[-i(\omega_1 t+\delta_1)]}\,,
~~v_n=g_n \exp{[-i(\omega_2 t+\delta_2)]}\,,
\ee
but now $f_n$ and $g_n$ satisfy
\be\label{m27}
f_n^2-ag_n^2=b\,,~~a,b>0\,.
\ee
On substituting this ansatz in Eqs. (\ref{m8}) and (\ref{m9}) we find that
this is a consistent ansatz provided
\be\label{m28}
a=-\frac{\mu_2}{\mu_1}\,,~~ b =-\frac{1}{\mu_1}\,,~~\omega_1 = \frac{\nu_1}{\mu_1}\,,
~~\omega_2= \frac{\nu_1 \mu_2}{\mu_1^2}\,.
\ee
This implies that such solutions are possible only if $\mu_1,\mu_2$ have 
opposite signs. 
Clearly this is a very general ansatz which admits a broad class of solutions.
As an illustration we discuss a few such solutions.
 
(vii) One solution is
\be\label{m29}
f_n= \frac{A}{\dn[\beta(n+c_2),m]}\,,~~ 
g_n=\frac{B \sqrt{m} \sn[\beta(n+c_2),m]}{\dn[\beta(n+c_2),m]}\,, 
\ee
provided Eq. (\ref{m28}) is satisfied and further
\be\label{m30}
b=A^2\,,~~\mu_1< 0\,,~~\mu_2>0\,,~~|\mu_1| A^2 = \mu_2 B^2\,.
\ee
In this case both $u_n,v_n$ satisfy the periodicity condition (\ref{m17a}). 
Note that if we interchange $f_n$ and $g_n$, then $\mu_1>0,\mu_2<0$.
In the limit $m=1$, this solution goes over to the hyperbolic solution
\be\label{m31a}
f_n=A \cosh[\beta(n+c_2)]\,,~~ g_n=B \sinh[\beta(n+c_2)]\,,
\ee

(viii) Another solution is
\be\label{m32}
f_n=\frac{A}{\dn^2[\beta(n+c_2),m]}+B\,,~~ 
g_n=\frac{F \sqrt{m} \sn[\beta(n+c_2),m]}{\dn^2[\beta(n+c_2),m]}\,,
\ee
provided Eq. (\ref{m28}) is satisfied and further
\be\label{m33}
b=\frac{A^2}{4}\,,~~\mu_1<0\,,~~\mu_2>0\,,~~|\mu_1| A^2 = \mu_2 F^2\,,
~~A=-2B\,.
\ee
In this case both $u_n,v_n$ satisfy the periodicity condition (\ref{m17a}). 
In the limit $m=1$, this solution goes over to the hyperbolic solution
\be\label{m31}
f_n=A \cosh^2[\beta(n+c_2)]\,,~~ g_n=B \sinh[\beta(n+c_2)]\cosh[\beta(n+c_2)]\,,
\ee 

(ix) Apart from these, several other solutions are possible. For example
one can have the following nonperiodic solution
\be\label{m34}
f_n = \frac{A\sqrt{(2+n^2)}}{\sqrt{1+n^2}}\,,~~g_n=\frac{B}{\sqrt{1+n^2}}\,,
\ee
provided Eq. (\ref{m28}) is satisfied and further
\be\label{m35}
b=A^2\,,~~\mu_1<0\,,~~\mu_2>0\,,~~|\mu_1| A^2 = \mu_2 F^2\,.
\ee
One can, easily  write down a wider class of such solutions.

\section{Solutions of the Coupled AL Model}

We show that for the coupled AL model characterized by Eqs. (\ref{m11}) and 
(\ref{m12}), one not only has solutions similar to those in the previous 
section 
(for the coupled Salerno case), but just like the uncoupled AL case
\cite{sb}, even 
coupled AL equations have moving periodic solutions in terms of Lam\'e 
polynomials of order one.

As in the previous section, if we start with the ansatz as given by 
Eqs. (\ref{m14}) and (\ref{m15}) or Eqs. (\ref{m26}) and (\ref{m27}), 
then it is easy to show that the entire 
discussion of the previous section goes through except that since in the
coupled AL model $\nu_1=2\mu_1,\nu_2=2\mu_2$, hence in the coupled AL model
with the above two ansatze, $\omega_1=2$,  $\omega_2=\frac{2\mu_2}{\mu_1}$. 
But for this minor change, all the nine solutions 
given in the previous section are also solutions of the coupled AL model
under the identical conditions (except $\omega_1=2, \omega_2 = 
\frac{2\mu_2}{\mu_1}$).

In addition to these nine solutions, we now show that as in the 
uncoupled case \cite{sb}, the coupled AL model also admits 
moving periodic solutions in terms of Lam\'e polynomials of order one.

(i) For example, it admits mixed moving periodic kink-pulse solution
\bea\label{al1}
&&u_n=A\exp[-i(\omega_1 t-k_1 n+\delta_1)] \dn[\beta(n-vt+\delta_2),m]\,, 
\nonumber \\
&&v_n=B\exp[-i(\omega_2 t-k_2 n+\delta_3)] \sqrt{m} 
\sn[\beta(n-vt+\delta_2),m]\,, 
\eea
provided
\bea\label{al2}
&&\omega_1=2\left[1-(1+\mu_2 B^2) \frac{\cos(k_1)\dn(\beta,m)}{\cn^2(\beta,m)}\right]\,, 
\nonumber \\
&&\omega_2=2\left[\frac{\mu_2}{\mu_1}-(1+\mu_2 B^2) 
\frac{\cos(k_2)\dn(\beta,m)}{\cn(\beta,m)}\right]\,, \nonumber \\
&&1=\mu_1 A^2\cs^2(\beta,m)-\mu_2 B^2\ns^2(\beta,m)\,, \nonumber \\
&&\beta v =\frac{2\sin(k_1)(1+\mu_2 B^2)}{\cs(\beta,m)}\,,
~~\cn(\beta,m)=\frac{\sin(k_2)}{\sin(k_1)}\,,
\eea
where $\cs(\beta,m)=\cn(\beta,m)/\sn(\beta,m)$ and 
$\ns(\beta,m)=1/\sn(\beta,m)$. 
Note that since it is a moving periodic kink-pulse solution, it must not only 
satisfy the periodicity condition (\ref{m17a}) but it must also satisfy
the periodicity condition
\be\label{al2z}
u_{n+\frac{2\pi}{k_1}}=u_n\,,~~v_{n+\frac{2\pi}{k_2}}=v_n\,.
\ee
The periodicity conditions (\ref{m17a}) and (\ref{al2z}) imply that $u_n,v_n$
are periodic solutions provided there exist integers $n_1,n_2,n_3,n_4$ such
that
\be\label{al2y}
n_1 \frac{2K(m)}{\beta} = n_2 \frac{2\pi}{k_1}\,,~~
n_3 \frac{4K(m)}{\beta} = n_4 \frac{2\pi}{k_2}\,.
\ee

(ii) Another mixed moving periodic kink-pulse solution that it admits is
\bea\label{al3}
&&u_n=A\exp[-i(\omega_1 t-k_1 n+\delta_1)] \sqrt{m} 
\cn[\beta(n-vt+\delta_2),m]\,, \nonumber \\
&&v_n=B\exp[-i(\omega_2 t-k_2 n+\delta_3)] \sqrt{m} 
\sn[\beta(n-vt+\delta_2),m]\,,
\eea
provided
\bea\label{al4}
&&\omega_1=2\left[1-(1+m\mu_2 B^2) \frac{\cos(k_1)\cn(\beta,m)}{\dn^2(\beta,m)}\right]\,, 
\nonumber \\
&&\omega_2=2\left[\frac{\mu_2}{\mu_1}-(1+m\mu_2 B^2) 
\frac{\cos(k_2)\cn(\beta,m)}{\dn(\beta,m)}\right]\,, 
\nonumber \\
&&1=\mu_1 A^2\ds^2(\beta,m)-\mu_2 B^2\ns^2(\beta,m)\,, \nonumber \\
&&\beta v =\frac{2\sin(k_1)}{\ds(\beta,m)}[1+m\mu_2 B^2]\,,
~~\dn(\beta,m)=\frac{\sin(k_2)}{\sin(k_1)}\,, 
\eea
where $\ds(\beta,m)=\dn(\beta,m)/\sn(\beta,m)$. 
Note that since it is a moving periodic kink-pulse solution, it must not only 
satisfy the periodicity condition (\ref{m19a}) but it must also satisfy
the periodicity condition (\ref{al2z}).
The periodicity conditions (\ref{m19a}) and (\ref{al2z}) imply that $u_n,v_n$
are periodic solutions provided there exist integers $n_1,n_2,n_3,n_4$ such
that
\be\label{al2x}
n_1 \frac{4K(m)}{\beta} = n_2 \frac{2\pi}{k_1}\,,~~
n_3 \frac{4K(m)}{\beta} = n_4 \frac{2\pi}{k_2}\,.
\ee

In the limit $m=1$, both these solutions reduce to the moving pulse-kink 
solution
\bea\label{al5}
&&u_n=A\exp[-i(\omega_1 t-k_1 n+\delta_1)] \sech[\beta(n-vt+\delta_2)]\,, 
\nonumber \\
&&v_n=B\exp[-i(\omega_2 t-k_2 n+\delta_3)] \tanh[\beta(n-vt+\delta_2)]\,,
\eea
provided
\bea\label{al6}
&&\sinh^2(\beta)=\mu_1 A^2-\mu_2 B^2 \cosh^2(\beta)\,, \nonumber \\
&&\omega_1=2[1-(1+\mu_2 B^2)\cos(k_1)\cosh(\beta)]\,,~~
\omega_2=2\left[\frac{\mu_2}{\mu_1}-(1+\mu_2 B^2)\cos(k_2)\right]\,, \nonumber \\
&&v\beta=2(1+\mu_2 B^2)\sin(k_1)\sinh(\beta)\,,~~
\frac{\sin(k_2)}{\sin(k_1)}= \sech (\beta)\,.
\eea
 
Notice that in case $k_1=k_2=v=0$,
the solutions (\ref{al1}), (\ref{al3}) and (\ref{al5}) become stationary 
coupled, periodic pulse-kink solutions provided relations (\ref{al2}),
(\ref{al4}) and (\ref{al6}) with $k_1=k_2=v=0$ are satisfied. However, we 
have already shown (and it can also be verified from relations (\ref{al2}),
(\ref{al4}) and (\ref{al6})) that the solutions (\ref{al1}), (\ref{al3}) 
and (\ref{al5}) with $k_1=k_2=v=0$
also hold good under the stronger conditions as given by Eqs. (\ref{m16}),
(\ref{m18}) and (\ref{m20}) with $\nu_1=2\mu_1$, $\nu_2=2\mu_2$. 

(iii) It also admits a coupled moving periodic pulse solution 
\bea\label{al7}
&&u_n=A\exp[-i(\omega_1 t-k_1 n+\delta_1)] \dn[\beta(n-vt+\delta_2),m]\,, 
\nonumber \\
&&v_n=B\exp[-i(\omega_2 t-k_2 n+\delta_3)] \dn[\beta(n-vt+\delta_2),m]\,, 
\eea
provided
\bea\label{al8}
&&k_1=k_2\,,~~\omega_1=2\left[1-\frac{\cos(k_1) \dn(\beta,m)}{\cn^2(\beta,m)}\right]\,,~~
\omega_2=2\left[\frac{\mu_2}{\mu_1}- \frac{\cos(k_1)\dn(\beta,m)}{\cn^2(\beta,m)}\right]
\,, \nonumber \\
&&1=(\mu_1 A^2+\mu_2 B^2)\cs^2(\beta,m)\,,~~
\beta v =\frac{2\sin(k_1)}{\cs(\beta,m)}\,.
\eea
Note that since it is a moving periodic pulse solution, it must not only 
satisfy the periodicity condition (\ref{m25z}) but it must also satisfy
the periodicity condition (\ref{al2z}).
The periodicity conditions (\ref{m25z}) and (\ref{al2z}) imply that $u_n,v_n$
are periodic solutions provided there exist integers $n_1,n_2,n_3,n_4$ such
that
\be\label{al2w}
n_1 \frac{2K(m)}{\beta} = n_2 \frac{2\pi}{k_1}\,,~~
n_3 \frac{2K(m)}{\beta} = n_4 \frac{2\pi}{k_2}\,.
\ee

(iv) Another coupled periodic moving pulse solution is
\bea\label{al9}
&&u_n=A\exp[-i(\omega_1 t-k_1 n+\delta_1)] \sqrt{m} 
\cn[\beta(n-vt+\delta_2),m]\,, \nonumber \\
&&v_n=B\exp[-i(\omega_2 t-k_2 n+\delta_3)] \sqrt{m} 
\cn[\beta(n-vt+\delta_2),m]\,, 
\eea
provided
\bea\label{al10}
&&k_1=k_2\,,~~\omega_1=2\left[1-\frac{\cos(k_1) \cn(\beta,m)}{\dn^2(\beta,m)}\right]\,,~~
\omega_2=2\left[\frac{\mu_2}{\mu_1}- \frac{\cos(k_1)\cn(\beta,m)}{\dn^2(\beta,m)}\right]
\,, \nonumber \\
&&1=(\mu_1 A^2+\mu_2 B^2)\ds^2(\beta,m)\,,~~
\beta v =\frac{2\sin(k_1)}{\ds(\beta,m)}\,.
\eea
Note that since it is a moving periodic pulse solution, it must not only 
satisfy the periodicity condition (\ref{m19a}) but it must also satisfy
the periodicity condition (\ref{al2z}).
The periodicity conditions (\ref{m19a}) and (\ref{al2z}) imply that $u_n,v_n$
are periodic solutions provided there exist integers $n_1,n_2,n_3,n_4$ such
that the condition (\ref{al2x}) is satisfied. 

(v) Finally, it also admits a mixed coupled moving periodic pulse solution
\bea\label{al11}
&&u_n=A\exp[-i(\omega_1 t-k_1 n+\delta_1)] \dn[\beta(n-vt+\delta_2),m]\,, 
\nonumber \\
&&v_n=B\exp[-i(\omega_2 t-k_2 n+\delta_3)] \sqrt{m} 
\cn[\beta(n-vt+\delta_2),m]\,, 
\eea
provided
\bea\label{al12}
&&\omega_1-2= -\left[1-(1-m)\mu_2 B^2\right]\frac{2\cos(k_1)
\dn(\beta,m)}{\cn^2(\beta,m)}\,,
~~\omega_2-2\frac{\mu_2}{\mu_1}= -\left[1-(1-m)\mu_2 B^2\right] 
\frac{2\cos(k_2)}{\cn(\beta,m)}\,, \nonumber \\
&&1=\mu_1 A^2 \cs^2(\beta,m)+\mu_2 B^2 \ds^2(\beta,m)\,,~~
\beta v =\frac{2\sin(k_1)}{\cs(\beta,m)}\left[1-(1-m)\mu_2 B^2\right]\,, \nonumber \\
&&\sin(k_1)\cn(\beta,m)=\sin(k_2) \dn(\beta,m)\,.
\eea
Note that since it is a moving periodic pulse solution, it must not only 
satisfy the periodicity condition (\ref{m17a}) but it must also satisfy
the periodicity condition (\ref{al2z}).
The periodicity conditions (\ref{m17a}) and (\ref{al2z}) imply that $u_n,v_n$
are periodic solutions provided there exist integers $n_1,n_2,n_3,n_4$ such
that the condition (\ref{al2y}) is satisfied. 

In the limit $m=1$, these three solution (iii), (iv), and (v) reduce to
\bea\label{al13}
&&u_n=A\exp[-i(\omega_1 t-k_1 n+\delta_1)] \sech[\beta(n-vt+\delta_2)]\,, 
\nonumber \\
&&v_n=B\exp[-i(\omega_2 t-k_2 n+\delta_3)] \sech[\beta(n-vt+\delta_2)]\,,
\eea
provided
\bea\label{al14}
&&k_1=k_2\,,~~\sinh^2(\beta)=(\mu_1 A^2+\mu_2 B^2)\,,  \nonumber \\
&&\omega_1=2[1-\cos(k_1)\cosh(\beta)]\,,~~
\omega_2=2\left[\frac{\mu_2}{\mu_1}-\cos(k_1)\cosh(\beta)\right]\,,~~
v\beta=2 \sin(k_1)\sinh(\beta)\,.
\eea
Thus the periodic pulse solutions exist provided at least one out of 
$\mu_1$, $\mu_2$ is positive.
For an entirely different coupled AL model, solution (\ref{al13}) has
also been obtained in \cite{yang}.

(vi) Finally, it also admits a coupled periodic kink solution
\bea\label{al15}
&&u_n=A\exp[-i(\omega_1 t-k_1 n +\delta_1)] \sqrt{m} 
\sn[\beta(n-vt+\delta_2),m]\,, \nonumber \\
&&v_n=B\exp[-i(\omega_2 t-k_2 n +\delta_3)] \sqrt{m} 
\sn[\beta(n-vt+\delta_2),m]\,,
\eea
provided
\bea\label{al16}
&&k_1=k_2\,,~~\omega_1=2[1-\cos(k_1)\cn(\beta,m)\dn(\beta,m)]\,, 
~~\omega_2=2\left[\frac{\mu_2}{\mu_1}-\cos(k_1)\cn(\beta,m)\dn(\beta,m)\right]\,, 
\nonumber \\
&&\mu_1 A^2+\mu_2 B^2= -\sn^2(\beta,m)\,,~~\beta v=2\sin(k_1)\sn(\beta,m)\,.
\eea
Thus this solution only exists if at least one out of $\mu_1$, $\mu_2$ is
negative.
Note that since it is a moving periodic kink solution, it must not only 
satisfy the periodicity condition (\ref{m19a}) but it must also satisfy
the periodicity condition (\ref{al2z}).
The periodicity conditions (\ref{m19a}) and (\ref{al2z}) imply that $u_n,v_n$
are periodic solutions provided there exist integers $n_1,n_2,n_3,n_4$ such
that the condition (\ref{al2x}) is satisfied. 

In the limit $m=1$, this solution reduces to
\bea\label{al17}
&&u_n=A\exp[-i(\omega_1 t-k_1 n+\delta_1)] \tanh[\beta(n-vt+\delta_2)]\,, 
\nonumber \\
&&v_n=B\exp[-i(\omega_2 t-k_2 n+\delta_3)] \tanh[\beta(n-vt+\delta_2)]\,,
\eea
provided
\bea\label{al18}
&&k_1=k_2\,,~~\omega_1=2[1-\cos(k_1) \sech^2(\beta)]\,, 
~~\omega_2=2\left[\frac{\mu_2}{\mu_1}-\cos(k_1) \sech^2(\beta)\right]\,, 
\nonumber \\
&&\mu_1 A^2+\mu_2 B^2= -\tanh^2(\beta)\,,~~\beta v=2\sin(k_1)\tanh(\beta)\,.
\eea

While obtaining these solutions, we have made use of several identities for
the Jacobi elliptic functions \cite{kls}.

\section{Solutions of the Coupled Saturated DNLS equations}
 
We show that unlike the coupled Salerno and the coupled AL case, the 
coupled saturated DNLS Eqs. (\ref{m5}) and (\ref{m6})
while they admit Lam\'e polynomial solutions of order one, they
{\it do not} admit general solutions characterized by Eqs. (\ref{m14}) and 
(\ref{m15}) or Eqs. (\ref{m26}) and (\ref{m27}). In particular, this 
model {\it does not} admit Lam\'e polynomial solutions of order two.
It is worth noting here that the uncoupled saturated DNLSE model does admit
Lam\'e polynomial solutions of order one \cite{krss}.

It is easy to check that the coupled Eqs. (\ref{m5}) and (\ref{m6}) have the 
following exact solutions in terms of Lam\'e polynomials of order one.

(i) It admits a coupled mixed pulse-kink solution
\bea\label{cs1}
&&u_n=A\exp[-i(\omega_1 t+\delta_1)] \dn[\beta(n+\delta_2),m]\,, \nonumber \\
&&v_n=B\exp[-i(\omega_2 t+\delta_3)] \sqrt{m} \sn[\beta(n+\delta_2),m]\,,
\eea
provided
\bea\label{cs2}
&&\omega_1=2-\frac{\nu_1}{\mu_1}\,, ~~\omega_2=2-\frac{\nu_2}{\mu_2}\,, 
~~\mu_2=\mu_1 \cn(\beta,m)\,,\nonumber \\
&&\mu_1 A^2 =\frac{\nu_1}{2\mu_1 \dn(\beta,m)}-1\,,~~
\mu_2 B^2 =\frac{\nu_1 \cn^2(\beta,m)}{2\mu_1 \dn(\beta,m)}-1\,.
\eea
For the solution (\ref{cs2}), $u_n,v_n$ satisfy the boundary condition
(\ref{m17a}).

(ii) Another coupled mixed pulse-kink solution is
\bea\label{cs3}
&&u_n=A\exp[-i(\omega_1 t+\delta_1)] \sqrt{m} \cn[\beta(n+\delta_2),m]\,, 
\nonumber \\
&&v_n=B\exp[-i(\omega_2 t+\delta_3)] \sqrt{m} \sn[\beta(n+\delta_2),m]\,,
\eea
provided
\bea\label{cs4}
&&\omega_1=2-\frac{\nu_1}{\mu_1}\,, ~~\omega_2=2-\frac{\nu_2}{\mu_2}\,, 
~~\mu_2=\mu_1 \dn(\beta,m)\,,\nonumber \\
&&m \mu_1 A^2 =\frac{\nu_1}{2\mu_1 \cn(\beta,m)}-1\,,~~
m \mu_2 B^2 =\frac{\nu_1 \dn^2(\beta,m)}{2\mu_1 \cn(\beta,m)}-1\,.
\eea
For the solution (\ref{cs3}), $u_n,v_n$ satisfy the boundary condition
(\ref{m19a}).

In the limit $m=1$, these two solutions (\ref{cs1})  and (\ref{cs3}) go over to
the mixed hyperbolic pulse-kink solution
\bea\label{cs5}
&&u_n=A\exp[-i(\omega_1 t+\delta_1)] \sech[\beta(n+\delta_2)]\,, \nonumber \\
&&v_n=B\exp[-i(\omega_2 t+\delta_3)] \tanh[\beta(n+\delta_2)]\,,
\eea
provided
\bea\label{cs6}
&&\omega_1=2-\frac{\nu_1}{\mu_1}\,, ~~\omega_2=2-\frac{\nu_2}{\mu_2}\,, 
~~\mu_2=\mu_1 \sech(\beta)\,,\nonumber \\
&&\mu_1 A^2 =\frac{\nu_1 \cosh(\beta)}{2\mu_1}-1\,,~~
 \mu_2 B^2 =\frac{\nu_1}{2\mu_1 \cosh(\beta)}-1\,.
\eea

(iii) This model also admits two coupled pulse solutions. One solution is
\bea\label{cs7}
&&u_n=A\exp[-i(\omega_1 t+\delta_1)] \dn[\beta(n+\delta_2),m]\,, \nonumber \\
&&v_n=B\exp[-i(\omega_2 t+\delta_3)] \dn[\beta(n+\delta_2),m]\,,
\eea
provided
\bea\label{cs8}
&&\omega_1=\omega_2=2\left[1-\frac{\dn(\beta,m)}{\cn^2(\beta,m)}\right]\,,~~
\nu_1=\nu_2\,,~~\mu_1=\mu_2\,, \nonumber \\
&&\mu_1(A^2+B^2)=\frac{\sn^2(\beta,m)}{\cn^2(\beta,m)}\,.
\eea
For the solution (\ref{cs7}), $u_n,v_n$ satisfy the boundary condition
(\ref{m25z}).

(iv) Another pulse solution is
\bea\label{cs9}
&&u_n=A\exp[-i(\omega_1 t+\delta_1)] \sqrt{m} \cn[\beta(n+\delta_2),m]\,, 
\nonumber \\
&&v_n=B\exp[-i(\omega_2 t+\delta_3)] \sqrt{m} \cn[\beta(n+\delta_2),m]\,,
\eea
provided
\bea\label{cs10}
&&\omega_1=\omega_2=2\left[1-\frac{\cn(\beta,m)}{\dn^2(\beta,m)}\right]\,,~~
\nu_1=\nu_2\,,~~\mu_1=\mu_2\,, \nonumber \\
&&\mu_1(A^2+B^2)=\frac{\sn^2(\beta,m)}{\dn^2(\beta,m)}\,.
\eea
For the solution (\ref{cs9}), $u_n,v_n$ satisfy the boundary condition
(\ref{m19a}).

In the limit $m=1$, these two  solutions (\ref{cs7}), (\ref{cs9}) 
reduce to the coupled hyperbolic pulse solution
\bea\label{cs11}
&&u_n=A\exp[-i(\omega_1 t+\delta_1)] \sech[\beta(n+\delta_2)]\,, \nonumber \\
&&v_n=B\exp[-i(\omega_2 t+\delta_3)] \sech[\beta(n+\delta_2)]\,,
\eea
provided
\bea\label{cs12}
&&\omega_1=\omega_2=-2[\cosh(\beta)-1]\,,~~
\nu_1=\nu_2\,,~~\mu_1=\mu_2\,, \nonumber \\
&&\mu_1(A^2+B^2)=\sinh^2(\beta)\,.
\eea
Notice that the coupled pulse solutions are admissible only if 
$\mu_1=\mu_2>0$.

(v) Finally, these coupled equations also admit a mixed kink solution 
\bea\label{cs13}
&&u_n=A\exp[-i(\omega_1 t+\delta_1)] \sqrt{m} \sn[\beta(n+\delta_2),m]\,, 
\nonumber \\
&&v_n=B\exp[-i(\omega_2 t+\delta_3)] \sqrt{m} \sn[\beta(n+\delta_2),m]\,,
\eea
provided
\bea\label{cs14}
&&\omega_1=\omega_2=2[1-\cn(\beta,m)\dn(\beta,m)]\,, \nonumber \\
&&\nu_1=\nu_2\,,~~\mu_1=\mu_2\,,~~\mu_1 (A^2+B^2)= -\sn^2(\beta,m)\,.
\eea
Thus unlike the coupled pulse solution, the coupled kink solution is only valid if $\mu_1=\mu_2 <0$.
For the solution (\ref{cs13}), $u_n,v_n$ satisfy the boundary condition
(\ref{m19a}).

In the limit $m=1$, this solution reduces to the hyperbolic kink solution
\bea\label{cs15}
&&u_n=A\exp[-i(\omega_1 t+\delta_1)] \tanh[\beta(n+\delta_2)]\,, \nonumber \\
&&v_n=B\exp[-i(\omega_2 t+\delta_3)] \tanh[\beta(n+\delta_2)]\,,
\eea
provided
\be\label{cs16}
\omega_1=\omega_2=2\tanh^2(\beta)\,,~~\mu_1=\mu_2\,,~~\nu_1=\nu_2\,,
~~\mu_1(A^2+B^2)=-\tanh^2(\beta)\,. 
\ee

Before ending this section, it might be worthwhile explaining why 
this model (unlike coupled Salerno or coupled AL models) does not admit 
Lam\'e polynomial solutions of order two. If we look at the field equations
which follow by using the general PB structure given by Eq. (\ref{m4}),
then it is easily seen that the model admits Lam\'e polynomials of order two   
as solutions provided
\be\label{csz}
1+\lambda_1 |u_n|^2 +\lambda_2 |v_n|^2 =0\,.
\ee
It is easily checked that while this condition can be readily satisfied in both
coupled Salerno and coupled AL models 
(where $\lambda_1 = \mu_1,\lambda_2=\mu_2$), this condition can never be 
satisfied in the coupled DNLS case since in that case 
$\lambda_1 = \lambda_2 =0$. Note, however, that in view of the nontrivial
identities for Jacobi elliptic functions \cite{kls} the coupled DNLS model 
still admits Lam\'e polynomial solutions of order one.  

\section{Solutions for a Coupled Discrete $\phi^6$ Model} 

We start from the same continuum coupled $\phi^6$ model for which recently 
we have obtained Lam\'e polynomial solutions of order two \cite{ak}. We 
now show that
if we consider the following discrete variant of the same model, then it has 
solutions not only in terms of Lam\'e polynomials of order one but even 
in terms of Lam\'e polynomials of order two, even though the Lam\'e 
polynomials of order two are {\it not} the solutions of the uncoupled
discrete $\phi^6$ model. 

The field equations for the static coupled continuum model, which we had 
considered recently \cite{ak}, are given by (modulo a factor of 2 in the
definitions of $c_1,c_2,e,f$) 
\be\label{4.1}
\frac{d^2\phi}{dx^2}=a_1 \phi-b_1\phi^3+2c_1\phi^5+d\phi \psi^2
+2e\phi^3 \psi^2 +2f\phi \psi^4\,,
\ee
\be\label{4.2}
\frac{d^2\psi}{dx^2}=a_2 \psi-b_2\psi^3+2c_2\psi^5+d\psi \phi^2
+e \phi^4 \psi+4f \phi^2 \psi^3\,.
\ee

Let us consider the following coupled discrete model
\be\label{4.3}
   \frac{1}{h^2} (\phi_{n+1}+\phi_{n-1}-2\phi_n)
   =a_1\phi_n-b_1 \phi_n^3+d \psi_n^2 \phi_n+[c_1 \phi_n^4+e\phi_n^2 \psi_n^2
+f\psi_n^4][\phi_{n+1}+\phi_{n-1}]\,, 
\ee
\be\label{4.4}
   \frac{1}{h^2} (\psi_{n+1}+\psi_{n-1}-2\psi_n)
   =a_2\psi_n- b_2 \psi_n^3+d \phi_n^2 \psi_n+[c_2 \psi_n^4+\frac{e}{2}\phi_n^4
+2f \phi_n^2 \psi_n^2][\psi_{n+1}+\psi_{n-1}]\,,
\ee
which in the continuum limit goes over to Eqs. (\ref{4.1}) and (\ref{4.2}).   
Here $h$ denotes the discreteness parameter. 

\noindent{\bf Solutions of the Uncoupled Model}

Let us first note that the uncoupled field Eq. (\ref{4.3}) for field $\phi$
(similar conclusion is also valid for the field $\psi$) given by
\be\label{4.5}
   \frac{1}{h^2} (\phi_{n+1}+\phi_{n-1}-2\phi_n)
   =a_1\phi_n-b_1 \phi_n^3+c_1 \phi_n^4 [\phi_{n+1}+\phi_{n-1}]\,, 
\ee
has three solutions in terms of Lam\'e polynomials of order one. However,  
Lam\'e
polynomials of order two {\it do not} satisfy the uncoupled Eq. (\ref{4.5}). 
In particular, it is easily shown that
\be\label{4.6}
\phi_n = A\dn[\beta(n+x_0),m]\,,
\ee
is an exact solution to the field Eq. (\ref{4.5}) provided
\be\label{4.7}
A^4 h^2 c_1 \cs^4(\beta,m) =1\,,~~a_1 =\frac{2}{h^2}
\left[\frac{\dn(\beta,m)}{\cn^2(\beta,m)}-1\right]\,,~~\frac{b_1^2}{2a_1 c_1}=
\frac{\dn^2(\beta,m)}{\cn^2(\beta,m)[\dn(\beta,m)-\cn^2(\beta,m)]}\,.
\ee
For the solution (\ref{4.6}), $\phi_n$ satisfies the boundary condition
\be\label{4.6z}
\phi_{n+\frac{2K(m)}{\beta}}=\phi_n\,.
\ee

Yet another solution to the field Eq. (\ref{4.5}) is given by
\be\label{4.8}
\phi_n = A\sqrt{m}\cn[\beta(n+x_0),m]\,,
\ee
provided
\be\label{4.9}
A^4 h^2 c_1 \ds^4(\beta,m) =1\,,~~a_1 =\frac{2}{h^2}
\left[\frac{\cn(\beta,m)}{\dn^2(\beta,m)}-1\right]\,,~~\frac{b_1^2}{2a_1 c_1}=
\frac{m^2\cn^2(\beta,m)}{\dn^2(\beta,m)[\cn(\beta,m)-\dn^2(\beta,m)]}\,.
\ee
For the solution (\ref{4.8}), $\phi_n$ satisfies the boundary condition
\be\label{4.8z}
\phi_{n+\frac{4K(m)}{\beta}}=\phi_n\,.
\ee

In the limit $m=1$, both these solutions go over to the pulse solution
\be\label{4.10}
\phi_n = A\sech[\beta(n+x_0)]\,,
\ee
provided
\be\label{4.11}
h^2 A^4 c_1 = \sinh^4(\beta)\,,~~a_1=\frac{2}{h^2}[\cosh(\beta)-1]>\,0\,,~~
\frac{b_1^2}{2a_1c_1}=\frac{\cosh^2(\beta)}{\cosh(\beta)-1}\,.
\ee

The third periodic solution to the field Eq. (\ref{4.5}) is given by
\be\label{4.12}
\phi_n = A\sqrt{m}\sn[\beta(n+x_0),m]\,,
\ee
provided
\be\label{4.13}
A^4 h^2 c_1 \ns^4(\beta,m) =1\,,~~a_1 =\frac{2}{h^2}
[\cn(\beta,m)\dn(\beta,m)-1] < 0\,,~~\frac{b_1^2}{2|a_1| c_1}=
\frac{m^2\cn^2(\beta,m)\dn^2(\beta,m)}{1-\cn(\beta,m)\dn(\beta,m)}\,.
\ee
For the solution (\ref{4.12}), $\phi_n$ satisfies the boundary condition
(\ref{4.8z}).

In the limit $m=1$, this solution goes over to the kink solution
\be\label{4.14}
\phi_n = A\tanh[\beta(n+x_0)]\,,
\ee
provided
\be\label{4.15}
h^2 A^4 c_1 = \tanh^4(\beta)\,,~~a_1=-\frac{2}{h^2}\tanh^2(\beta)<\,0\,,~~
\frac{b_1^2}{2|a_1|c_1}=\frac{4}{\sinh^2(2\beta)}\,.
\ee

Let us now discuss the solutions of the coupled Eqs. (\ref{4.3}) 
and (\ref{4.4}). It turns out that as in the coupled AL case, the 
$\phi^6$ coupled equations have solutions satisfying the ansatz 
similar to  (\ref{m15}) (but no solutions satisfying the ansatz similar to 
(\ref{m27})), and also solutions in terms of Lam\'e polynomials of order
one (by making use of the identities for the Jacobi elliptic functions \cite{kls}).

\noindent{\bf Solutions of the Coupled Model Satisfying Ansatz Similar to (\ref{m15})}

On substituting the ansatz  
\be\label{4.16z}
\phi_n^2+a\psi_n^2 =b\,,~~a,b\, > \,0\,,
\ee
(which is similar to the ansatz (\ref{m15})) in
the coupled field Eqs. (\ref{4.3}) and (\ref{4.4}), we find that such
solutions exist provided
\bea\label{4.16a}
&&c_1=c_2=f=\frac{e}{2}\,,~~b_1=b_2=-d\,,~~a_1=a_2\,,~~c_1 h^2 b^2 =1\,,
\nonumber\\ 
&&a=1\,,~~a_1+\frac{2}{h^2}=\frac{b_1}{\sqrt{h^2c_1}}\,.
\eea
This is a rather general ansatz and there are several solutions of this type
which exist for this model.

\noindent{\bf Lam\'e polynomial solutions of order one}

(i) One solution is

\be\label{4.17a}
\phi_n=A \dn[\beta(n+c_2),m]\,,~~ \psi_n=B \sqrt{m} \sn[\beta(n+c_2),m]\,,
\ee
provided Eq. (\ref{4.16a}) is satisfied and further
\be\label{4.18a}
b=A^2\,,~~A^2 = B^2\,,~~a_1-b_1 A^2+2c_1 A^4=0\,.
\ee
Note that the width $\beta$ is completely arbitrary.
For this solution, $\phi_n,\psi_n$ satisfy the boundary condition 
given by Eq. (\ref{m17a}) with $\phi_n,\psi_n$ replacing $f_n,g_n$
respectively.

(ii) Another solution is

\be\label{4.19a}
\phi_n=A \sqrt{m} \cn[\beta(n+c_2),m]\,,~~ \psi_n=B \sqrt{m} \sn[\beta(n+c_2),m]\,,
\ee
provided Eq. (\ref{4.16a}) is satisfied and further
\be\label{4.20a}
b=m A^2\,,~~A^2 = B^2\,, ~~ a_1-mb_1 A^2+2m^2 c_1 A^4 = 0\,.
\ee
For this solution, $\phi_n,\psi_n$ satisfy the boundary condition 
given by Eq. (\ref{m19a}) with $\phi_n,\psi_n$ replacing $f_n,g_n$
respectively.

In the limit $m=1$, both these solutions go over to the hyperbolic solution
\be\label{4.21a}
f_n=A \sech[\beta(n+c_2)]\,,~~ g_n=B \tanh[\beta(n+c_2)]\,,
\ee

\noindent{\bf Lam\'e polynomial solutions of order two}

(iii) One solution is given by
\be\label{4.22a}
\phi_n=A \dn^2[\beta(n+c_2),m]+B\,,~~ 
\psi_n=F \sqrt{m} \sn[\beta(n+c_2),m] \dn[\beta(n+c_2),m]\,,
\ee
provided Eq. (\ref{4.16a}) is satisfied and further
\be\label{4.23a}
b=\frac{A^2}{4}\,, ~~A^2 =  F^2\,,~~A=-2B\,,~~a_1-b_1B^2+2c_1B^4 =0\,.
\ee
For this solution, $\phi_n,\psi_n$ satisfy the boundary condition 
given by Eq. (\ref{m17a}) with $\phi_n,\psi_n$ replacing $f_n,g_n$
respectively.

(iv) Another solution is

\be\label{4.24a}
\phi_n=A \dn^2[\beta(n+c_2),m]+B\,,~~ 
\psi_n=F m \sn[\beta(n+c_2),m] \cn[\beta(n+c_2),m]\,,
\ee
provided Eq. (\ref{4.16a}) is satisfied and further
\be\label{4.25a}
b=\frac{m^2 A^2}{4}\,,~~A^2 = F^2\,,~~(2-m)A=-2B\,,
~~8a_1-2b_1 m^2 F^2 +c_1 m^4 F^4 =0\,.
\ee
For this solution, $\phi_n,\psi_n$ satisfy the boundary condition 
given by Eq. (\ref{m25z}) with $\phi_n,\psi_n$ replacing $f_n,g_n$
respectively.

In the limit $m=1$, both the solutions (\ref{4.22a}) and (\ref{4.24a}), 
go over to the hyperbolic solution
\be\label{4.26a}
\phi_n=A \sech^2[\beta(n+c_2)]+B\,,~~ 
\psi_n=F \tanh[\beta(n+c_2)] \sech[\beta(n+c_2)]\,.
\ee

(v) Apart from these, several other solutions are possible. For example
one can have the following nonperiodic solution
\be\label{4.27a}
\phi_n = \frac{A}{\sqrt{1+n^2}}\,,~~\psi_n=\frac{Bn}{\sqrt{1+n^2}}\,,
\ee
provided Eq. (\ref{4.16a}) is satisfied and further
\be\label{4.28a}
b=A^2\,,~~A^2 = B^2\,.
\ee

(vi) Yet another solution is
\be\label{4.29a}
\phi_n=A \cos[\beta(n+c_2)]\,,~~ \psi_n=B \sin[\beta(n+c_2)]\,,
\ee
provided Eq. (\ref{4.16a}) is satisfied and further 
\be\label{4.30a}
b=A^2\,, ~~A^2 =  B^2\,.
\ee
For this solution, $\phi_n,\psi_n$ satisfy the boundary condition 
given by Eq. (\ref{m25cz}) with $\phi_n,\psi_n$ replacing $f_n,g_n$
respectively.

\noindent{\bf Solutions following from identities for Jacobi elliptic functions}

Using the identities for the Jacobi elliptic functions \cite {kls}, we now 
show that there are six Lam\'e polynomial solutions of order one to the
coupled field Eqs. (\ref{4.3}) and (\ref{4.4}).

\noindent{\bf Solution 1}: It is not difficult to show that
\be\label{4.16}
\phi_n = A\dn[\beta(n+x_o),m]\,,~~\psi_n=B\sqrt{m}\sn[\beta(n+x_0),m]\,,
\ee
is an  exact solution to the coupled field Eqs. (\ref{4.3}) and (\ref{4.4})
provided
\be\label{4.17}
b_1A^2+dB^2=2(c_1A^4+fB^4-eA^2B^2)\ds(\beta,m)\ns(\beta,m)\,,
\ee
\be\label{4.18}
a_1+\frac{2}{h^2}+dB^2+2(eA^2B^2-2fB^4)\ds(\beta,m)\ns(\beta,m)=
2(c_1A^4+fB^4-eA^2B^2)\ds(\beta,m)\ns(\beta,m)\cs^2(\beta,m)\,,
\ee
\be\label{4.19}
\frac{1}{h^2}-fB^4+(eA^2B^2-2fB^4)\cs^2(\beta,m)=
(c_1A^4+fB^4-eA^2B^2)\cs^4(\beta,m)\,,
\ee
\be\label{4.20}
b_2B^2+dA^2=-2\left(c_2B^4+\frac{e}{2}A^4-2fA^2B^2\right)
\ds(\beta,m)\cs(\beta,m)\,,
\ee
\be\label{4.21}
a_2+\frac{2}{h^2}+dA^2-2(2fA^2B^2-eA^4)\ds(\beta,m)\cs(\beta,m)=
2(c_2B^4+\frac{e}{2}A^4-2fA^2B^2)\ds(\beta,m)\cs(\beta,m)\ns^2(\beta,m)\,,
\ee
\be\label{4.22}
\frac{1}{h^2}-\frac{e}{2}A^4-(2fA^2B^2-eA^4)\ns^2(\beta,m)=
\left(c_2B^4+\frac{e}{2}A^4-2fA^2B^2\right)\ns^4(\beta,m)\,.
\ee
For this solution, $\phi_n,\psi_n$ satisfy the boundary condition 
given by Eq. (\ref{m17a}) with $\phi_n,\psi_n$ replacing $f_n,g_n$
respectively.

These equations are of course trivially satisfied if relations (\ref{4.16a})
and (\ref{4.18a}) are satisfied.
It is worth pointing out that while relations (\ref{4.16a}) and (\ref{4.18a})
are sufficient so that (\ref{4.16}) constitutes an exact solution to the 
coupled Eqs. (\ref{4.3}) and (\ref{4.4}), it is not obvious if relations 
(\ref{4.16a}) and (\ref{4.18a}) are also necessary. The necessary relations 
are as given by Eqs. (\ref{4.17}) to (\ref{4.22}).  

\noindent{\bf Solution 2}: Another solution is given by
\be\label{4.24}
\phi_n = A\sqrt{m}\cn[\beta(n+x_o),m]\,,~~\psi_n=B\sqrt{m}\sn[\beta(n+x_0),m]\,,
\ee
which is an exact solution to the coupled field Eqs. (\ref{4.3}) and (\ref{4.4})
provided
\be\label{4.25}
b_1A^2+dB^2=2(c_1A^4+fB^4-eA^2B^2)\cs(\beta,m)\ns(\beta,m)\,,
\ee
\be\label{4.26}
a_1+\frac{2}{h^2}+mdB^2+2m(eA^2B^2-2fB^4)\cs(\beta,m)\ns(\beta,m)=
2(c_1A^4+fB^4-eA^2B^2)\cs(\beta,m)\ns(\beta,m)\ds^2(\beta,m)\,,
\ee
\be\label{4.27}
\frac{1}{h^2}-m^2fB^4+m(eA^2B^2-2fB^4)\ds^2(\beta,m)=
(c_1A^4+fB^4-eA^2B^2)\ds^4(\beta,m)\,,
\ee
\be\label{4.28}
b_2B^2+dA^2=-2\left(c_2B^4+\frac{e}{2}A^4-2fA^2B^2\right)
\ds(\beta,m)\cs(\beta,m)\,,
\ee
\be\label{4.29}
a_2+\frac{2}{h^2}+mdA^2-2(2fA^2B^2-eA^4)\cs(\beta,m)\cs(\beta,m)=
2\left(c_2B^4+\frac{e}{2}A^4-2fA^2B^2\right)
\ds(\beta,m)\cs(\beta,m)\ns^2(\beta,m)\,,
\ee
\be\label{4.30}
\frac{1}{h^2}-\frac{e}{2}m^2A^4-(2fA^2B^2-eA^4)\ns^2(\beta,m)=
\left(c_2B^4+\frac{e}{2}A^4-2fA^2B^2\right)\ns^4(\beta,m)\,.
\ee
For this solution, $\phi_n,\psi_n$ satisfy the boundary condition 
given by Eq. (\ref{m19a}) with $\phi_n,\psi_n$ replacing $f_n,g_n$
respectively.

These equations are trivially satisfied if relations (\ref{4.16a}) and 
(\ref{4.20a}) are satisfied.

In the limit $m=1$, both the solutions (\ref{4.16}) and (\ref{4.24}) go over to 
the hyperbolic soliton solution 
\be\label{4.32}
\phi_n=A\sech[\beta(n+x_0)]\,,~~\psi_n=B\tanh[\beta(n+x_0)]\,,
\ee
provided relations (\ref{4.17}) to (\ref{4.22}) with $m=1$ are satisfied. 

\noindent{\bf Solution 3}: It is not difficult to show that
\be\label{4.33}
\phi_n = A\dn[\beta(n+x_o),m]\,,~~\psi_n=B\sqrt{m}\cn[\beta(n+x_0),m]\,,
\ee
is an  exact solution to the coupled field Eqs. (\ref{4.3}) and (\ref{4.4})
provided
\be\label{4.34}
b_1A^2-dB^2=2(c_1A^4+fB^4+eA^2B^2)\ds(\beta,m)\ns(\beta,m)\,,
\ee
\bea\label{4.35}
&&a_1+\frac{2}{h^2}-(1-m)dB^2-2(1-m)(eA^2B^2+2fB^4)
\ds(\beta,m)\ns(\beta,m) \nonumber \\
&&=2(c_1A^4+fB^4+eA^2B^2)\ds(\beta,m)\ns(\beta,m)\cs^2(\beta,m)\,,
\eea
\be\label{4.36}
\frac{1}{h^2}-(1-m)^2fB^4-(1-m)(eA^2B^2+2fB^4)\cs^2(\beta,m)=
(c_1A^4+fB^4+eA^2B^2)\cs^4(\beta,m)\,,
\ee
\be\label{4.37}
b_2B^2-dA^2=-2\left(c_2B^4+\frac{e}{2}A^4+2fA^2B^2\right)
\ns(\beta,m)\cs(\beta,m)\,,
\ee
\bea\label{4.38}
&&a_2+\frac{2}{h^2}+(1-m)dA^2+2(1-m)(2fA^2B^2+eA^4)\ns(\beta,m)\cs(\beta,m) 
\nonumber \\
&&=2\left(c_2B^4+\frac{e}{2}A^4+2fA^2B^2\right)
\ns(\beta,m)\cs(\beta,m)\ds^2(\beta,m)\,,
\eea
\be\label{4.39}
\frac{1}{h^2}-\frac{e}{2}(1-m)^2 A^4+(1-m)(2fA^2B^2+eA^4)\ds^2(\beta,m)=
(c_2B^4+\frac{e}{2}A^4+2fA^2B^2)\ds^4(\beta,m)\,.
\ee
For this solution, $\phi_n,\psi_n$ satisfy the boundary condition 
given by Eq. (\ref{m17a}) with $\phi_n,\psi_n$ replacing $f_n,g_n$
respectively.

\noindent{\bf Solution 4}: Another solution to the coupled Eqs. (\ref{4.3}) and
(\ref{4.4}) is given by
\be\label{4.40}
\phi_n = A\dn[\beta(n+x_o),m]\,,~~\psi_n=B\dn[\beta(n+x_0),m]\,,
\ee
provided
\be\label{4.41}
b_1A^2-dB^2=2(c_1A^4+fB^4+eA^2B^2)\ds(\beta,m)\ns(\beta,m)\,,
\ee
\be\label{4.42}
a_1+\frac{2}{h^2}
=2(c_1A^4+fB^4+eA^2B^2)\ds(D,m)\ns(\beta,m)\cs^2(\beta,m)\,,
\ee
\be\label{4.43}
\frac{1}{h^2}=(c_1A^4+fB^4+eA^2B^2)\cs^4(\beta,m)\,,
\ee
\be\label{4.44}
b_2B^2-dA^2=2\left(c_2B^4+\frac{e}{2}A^4+2fA^2B^2\right)
\ds(\beta,m)\ns(\beta,m)\,,
\ee
\be\label{4.45}
a_2+\frac{2}{h^2} =2\left(c_2B^4+\frac{e}{2}A^4
+2fA^2B^2\right)\ns(\beta,m)\ds(\beta,m)\cs^2(\beta,m)\,,
\ee
\be\label{4.46}
\frac{1}{h^2}=\left(c_2B^4+\frac{e}{2}A^4+2fA^2B^2\right)\cs^4(\beta,m)\,.
\ee
For this solution, $\phi_n,\psi_n$ satisfy the boundary condition 
given by Eq. (\ref{m25z}) with $\phi_n,\psi_n$ replacing $f_n,g_n$
respectively.

\noindent{\bf Solution 5}: Yet another solution to the coupled Eqs. (\ref{4.3}) and
(\ref{4.4}) is given by
\be\label{4.47}
\phi_n = A\sqrt{m}\cn[\beta(n+x_o),m]\,,~~\psi_n=B\sqrt{m}\cn[\beta(n+x_0),m]\,,
\ee
provided
\be\label{4.48}
b_1A^2-dB^2=2(c_1A^4+fB^4+eA^2B^2)\cs(\beta,m)\ns(\beta,m)\,,
\ee
\be\label{4.49}
a_1+\frac{2}{h^2}
=2(c_1A^4+fB^4+eA^2B^2)\cs(\beta,m)\ns(\beta,m)\ds^2(\beta,m)\,,
\ee
\be\label{4.50}
\frac{1}{h^2}=(c_1A^4+fB^4+eA^2B^2)\ds^4(\beta,m)\,,
\ee
\be\label{4.51}
b_2B^2-dA^2=2\left(c_2B^4+\frac{e}{2}A^4+2fA^2B^2\right) 
\cs(\beta,m)\ns(\beta,m)\,,
\ee
\be\label{4.52}
a_2+\frac{2}{h^2} =2\left(c_2B^4+\frac{e}{2}A^4
+2fA^2B^2\right)\ns(\beta,m)\cs(\beta,m)\ds^2(\beta,m)\,,
\ee
\be\label{4.53}
\frac{1}{h^2}=\left(c_2B^4+\frac{e}{2}A^4+2fA^2B^2\right)\ds^4(\beta,m)\,.
\ee
For this solution, $\phi_n,\psi_n$ satisfy the boundary condition 
given by Eq. (\ref{m19a}) with $\phi_n,\psi_n$ replacing $f_n,g_n$
respectively.

In the limit $m=1$, all three solutions given by (\ref{4.33}), (\ref{4.40})
and (\ref{4.47})  go over to the hyperbolic soliton solution
\be\label{4.54}
\phi_n=A\sech[\beta(n+x_0)]\,,~~\psi_n=B\sech[\beta(n+x_0)]\,,
\ee
provided
\be\label{4.55}
a_1=a_2=\frac{2}{h^2}[\cosh(\beta)-1]>\,0\,,~~b_1A^2-dB^2=b_2B^2-dA^2=
\frac{2}{h^2}\sinh^2(\beta)\cosh(\beta)\,,
\ee
\be\label{4.56}
c_1A^4+eA^2B^2+fB^4=c_2 B^4+2fA^2B^2+\frac{e}{2}A^4
=\frac{\sinh^4(\beta)}{h^2}\,.
\ee

\noindent{\bf Solution 6}: Finally, another solution to the coupled Eqs. (\ref{4.3}) and
(\ref{4.4}) is given by
\be\label{4.57}
\phi_n = A\sqrt{m}\sn[\beta(n+x_o),m]\,,~~\psi_n=B\sqrt{m}\sn[\beta(n+x_0),m]\,,
\ee
provided
\be\label{4.58}
b_1A^2-dB^2=2(c_1A^4+fB^4+eA^2B^2)\cs(\beta,m)\ds(\beta,m)\,,
\ee
\be\label{4.59}
a_1+\frac{2}{h^2}
=2(c_1A^4+fB^4+eA^2B^2)\cs(\beta,m)\ds(\beta,m)\ns^2(\beta,m)\,,
\ee
\be\label{4.60}
\frac{1}{h^2}=(c_1A^4+fB^4+eA^2B^2)\ns^4(\beta,m)\,,
\ee
\be\label{4.61}
b_2B^2-dA^2=2(c_2B^4+\frac{e}{2}A^4+2fA^2B^2)\cs(\beta,m)\ds(\beta,m)\,,
\ee
\be\label{4.62}
a_2+\frac{2}{h^2} =2(c_2B^4+\frac{e}{2}A^4
+2fA^2B^2)\ds(\beta,m)\cs(\beta,m)\ns^2(\beta,m)\,,
\ee
\be\label{4.63}
\frac{1}{h^2}=(c_2B^4+\frac{e}{2}A^4+2fA^2B^2)\ns^4(\beta,m)\,.
\ee
For this solution, $\phi_n,\psi_n$ satisfy the boundary condition 
given by Eq. (\ref{m19a}) with $\phi_n,\psi_n$ replacing $f_n,g_n$
respectively.

In the limit $m=1$, this solution goes over to the 
hyperbolic soliton solution
\be\label{4.64}
\phi_n=A\tanh[\beta(n+x_0)]\,,~~\psi_n=B\tanh[\beta(n+x_0)]\,,
\ee
provided
\be\label{4.65}
a_1=a_2=-\frac{2}{h^2}[\tanh^2(\beta)]<\,0\,,~~b_1A^2-dB^2=b_2B^2-dA^2=
-\frac{2}{h^2}\frac{\sinh^2(\beta)}{\cosh^4(\beta)}\,,
\ee
\be\label{4.66}
c_1A^4+eA^2B^2+fB^4=c_2 B^4+2fA^2B^2+\frac{e}{2}A^4
=\frac{\tanh^4(\beta)}{h^2}\,.
\ee

\section{Solutions for a Coupled Discrete $\phi^4$ Model} 

We start from the same coupled static discrete field equations as in our recent 
paper \cite{ak1} for which we had obtained six solutions in terms of Lam\'e 
polynomials of order one. We now show that the same model also admits
two Lam\'e polynomial solutions of order two, even though they are {\it not}
the solutions of the corresponding uncoupled problem. Let us start from the 
field equations considered in \cite{ak1}
\be\label{5.1}
   \frac{1}{h^2} (\phi_{n+1}+\phi_{n-1}-2\phi_n)
   -2\alpha_1\phi_n-[2\beta_1\phi_n^2+\gamma \psi_n^2]
   [\phi_{n+1}+\phi_{n-1}]=0\,, 
\ee
\be\label{5.2}
   \frac{1}{h^2} (\psi_{n+1}+\psi_{n-1}-2\psi_n)
   -2\alpha_2\psi_n-[2\beta_2\psi_n^2+\gamma \phi_n^2]
   [\psi_{n+1}+\psi_{n-1}]=0\,.
\ee

Let us now discuss the solutions of the coupled Eqs. (\ref{5.1}) 
and (\ref{5.2}). It turns out that as in the coupled $\phi^6$ case, the 
$\phi^4$ coupled equations have solutions satisfying ansatz similar to
the one given by Eq. (\ref{m15}) (but no solutions satisfying ansatz  
similar to (\ref{m27})). Further, they also have solutions in terms of 
Lam\'e polynomials 
of order one which we have already discussed in \cite{ak1}. Note that these 
solutions were obtained by making use of the identities for the Jacobi elliptic 
functions \cite{kls}.

\noindent{\bf Solutions satisfying ansatz similar to (\ref{m15})}

On substituting the ansatz as given by Eq. (\ref{4.16z}) (which is similar
to the ansatz given by Eq. (\ref{m15})) in
the coupled field Eqs. (\ref{5.1}) and (\ref{5.2}), we find that such
solutions exist provided
\be\label{5.3}
2\beta_1=2\beta_2=\gamma\,,~~\alpha_1=\alpha_2=-\frac{1}{h^2}\,,~~a=1\,.
\ee
This is a rather general ansatz and there are several solutions of this type
which exist in this model.

\noindent{\bf Lam\'e polynomial solutions of order one}

(i) One solution is

\be\label{5.4}
\phi_n=A \dn[\beta(n+c_2),m]\,,~~ \psi_n=B \sqrt{m} \sn[\beta(n+c_2),m]\,,
\ee
provided Eq. (\ref{5.3}) is satisfied and further
\be\label{5.5}
b=A^2=\frac{1}{2\beta_1 h^2}\,,~~A^2 = B^2\,.
\ee
Note that the width $\beta$ is completely arbitrary.
For this solution, $\phi_n,\psi_n$ satisfy the boundary condition 
given by Eq. (\ref{m17a}) with $\phi_n,\psi_n$ replacing $f_n,g_n$
respectively.

(ii) Another solution is

\be\label{5.6}
\phi_n=A \sqrt{m} \cn[\beta(n+c_2),m]\,,~~ 
\psi_n=B \sqrt{m} \sn[\beta(n+c_2),m]\,,
\ee
provided Eq. (\ref{5.3}) is satisfied and further
\be\label{5.7}
b=m A^2 = \frac{1}{2\beta_1 h^2}\,,~~A^2 = B^2\,.
\ee
For this solution, $\phi_n,\psi_n$ satisfy the boundary condition 
given by Eq. (\ref{m19a}) with $\phi_n,\psi_n$ replacing $f_n,g_n$
respectively.

In the limit $m=1$, both these solutions go over to the hyperbolic solution
\be\label{5.8}
\phi_n=A \sech[\beta(n+c_2)]\,,~~ \psi_n=B \tanh[\beta(n+c_2)]\,,
\ee

\noindent{\bf Lam\'e polynomial solutions of order two}

(iii) One solution is given by
\be\label{5.9}
\phi_n=A \dn^2[\beta(n+c_2),m]+B\,,~~ 
\psi_n=F \sqrt{m} \sn[\beta(n+c_2),m] \dn[\beta(n+c_2),m]\,,
\ee
provided Eq. (\ref{5.3}) is satisfied and further
\be\label{5.10}
b=\frac{A^2}{4}=\frac{1}{2\beta_1 h^2}\,, ~~A^2 =  F^2\,,~~A=-2B\,.
\ee
For this solution, $\phi_n,\psi_n$ satisfy the boundary condition 
given by Eq. (\ref{m19a}) with $\phi_n,\psi_n$ replacing $f_n,g_n$
respectively.

(iv) Another solution is

\be\label{5.11}
\phi_n=A \dn^2[\beta(n+c_2),m]+B\,,~~ 
\psi_n=F m \sn[\beta(n+c_2),m] \cn[\beta(n+c_2),m]\,,
\ee
provided Eq. (\ref{5.3}) is satisfied and further
\be\label{5.12}
b=\frac{m^2 A^2}{4}\,,~~A^2 = F^2\,,~~(2-m)A=-2B\,.
\ee
For this solution, $\phi_n,\psi_n$ satisfy the boundary condition 
given by Eq. (\ref{m25z}) with $\phi_n,\psi_n$ replacing $f_n,g_n$
respectively.

In the limit $m=1$, both solutions (\ref{5.9}) and (\ref{5.11}) go over to 
the hyperbolic solution
\be\label{5.13}
\phi_n=A \sech^2[\beta(n+c_2)]+B\,,~~ 
\psi_n=F \tanh[\beta(n+c_2)] \sech[\beta(n+c_2)]\,.
\ee

(v) Apart from these, several other solutions are possible. For example
one can have a nonperiodic solution
\be\label{5.14}
\phi_n = \frac{A}{\sqrt{1+n^2}}\,,~~\psi_n=\frac{Bn}{\sqrt{1+n^2}}\,,
\ee
provided Eq. (\ref{5.3}) is satisfied and further
\be\label{5.15}
b=A^2\,,~~A^2 = B^2\,.
\ee

(vi) Yet another solution is
\be\label{5.16}
\phi_n=A \cos[\beta(n+c_2)]\,,~~ \psi_n=B \sin[\beta(n+c_2)]\,,
\ee
provided Eq. (\ref{5.3}) is satisfied and further 
\be\label{5.17}
b=A^2\,, ~~A^2 =  B^2\,.
\ee
For this solution, $\phi_n,\psi_n$ satisfy the boundary condition 
given by Eq. (\ref{m25z}) with $\phi_n,\psi_n$ replacing $f_n,g_n$
respectively.

\section{Summary}

In this paper we have shown that for a number of {\it coupled discrete} models,
e.g. coupled Salerno, coupled Ablowitz-Ladik, coupled saturated nonlinear 
Schr\"odinger equation, coupled $\phi^6$, coupled $\phi^4$, while the uncoupled 
equations do not admit solutions in terms of Lam\'e polynomials of order two, 
the coupled models do admit such solutions.  These solutions (with appropriate 
boundary conditions) have relevance in physical contexts ranging from 
ferroelectric [7, 8, 9] to multiferroic [4, 5, 6] materials to the models in field 
theory [10, 14] as well as for various discrete contexts [11, 12, 13]. 

The stability of various solutions found here remains an open issue to be 
explored numerically, particularly some solutions have an arbitrary soliton width.  
In addition, the scattering of solitons of various discrete models is an important 
issue with these static solutions boosted with a certain velocity. Similarly, the 
Peierls-Nabarro (discreteness) barrier for the solutions remains to be explored.  
Given the solutions in terms of Lam\'e functions of order one and two, it is then 
worth enquiring if one considers coupling of three discrete fields, would they 
admit solutions in terms of Lam\'e polynomials of order three?  And if true, can 
one generalize it to the case of $N$ coupled fields? We hope to address these 
issues in the near future.

\section{Acknowledgment}
A.K. acknowledges the hospitality of the Center for Nonlinear
studies at LANL.  This work was supported in part by the U.S.
Department of Energy.

\end{document}